\documentclass[11pt]{article}
\usepackage[centertags]{amsmath}
\usepackage{amsfonts}
\usepackage{amssymb}
\usepackage{amsthm}
\usepackage{newlfont}
\usepackage{graphicx}

\newfont{\bb}{msbm10 at 12pt}

\begin{document}

\title{Conformal Ricci Collineations of Plane Symmetric Static Spacetimes}
\author{ Ahmad T Ali\\
King Abdulaziz University,\\
Faculty of Science, Department of Mathematics,\\
PO Box 80203, Jeddah, 21589, Saudi Arabia.\\
Mathematics Department,\\
Faculty of Science, Al-Azhar University,\\
Nasr City, 11884, Cairo, Egypt.\\
E-mail: ahmadtawfik@gmail.com@yahoo.com}

%\date{}

\maketitle
\begin{abstract} This article explores the Conformal Ricci Collineations (CRCs) for the plane-symmetric static
spacetime. The non-linear coupled CRC equations are solved to get
the general form of conformal Ricci symmetries. In the
non-degenerate case, it turns out that the dimension of the Lie
algebra of CRCs is finite. In the case were the Ricci tensor is
degenerate, it found that the algebra of CRCs for the
plane-symmetric static spacetime is mostly, but not always, infinite
dimensional. In one case of degenerate Ricci tensor, we solved the
differential constraints completely and a spacetime metric is
obtained along with CRCs. We found ten possible cases of finite and
infinite dimensional Lie algebras of CRCs for the considered
spacetime.
\end{abstract}

%\emph{PACS:} ***.


%%%%%%%%%%%%%%%%%%%%%%%%%%%%%%%%%%%%%%%%%%%%%%%
\section{Introduction }
%%%%%%%%%%%%%%%%%%%%%%%%%%%%%%%%%%%%%%%%%%%%%%%

The general theory of relativity proposed by Einstien is an elegant
theory of gravitation which has provided an understanding of the
mystery of gravity at least at the classical level. An important
benefit of the general theory of relativity is that it helps us
understand the large-scale structure of the universe. In theoretical
physics one has two main tools to study the properties of evolution
of dynamical systems: Symmetries of the equations of motion and
Collineations (symmetries) of the spacetime. The symmetries
(symmetry analysis method) of differential equations (ordinary and
partial) are a powerful method to find the exact (invariant)
solutions of Einstein field equations \cite{ali1, ali2, att1, olve1,
ovsi1, yadav1}. The symmetries of the spacetime play a central role
in the classification of the exact solutions of the Einstein field
equations. The classification of spacetimes according to different
types of symmetries is an important part of recent research in the
field of general relativity \cite{hall1, steph1}. The most basic
symmetries of a spacetimes are the \textit{Killing symmetries} or
\textit{isometries}. Killing symmetries of a spacetimes are vector
fields along which the metric tensor remains invariant under the Lie
transport. The components of Killing vector fields satisfy the
\textit{Killing equation} which is given as
\begin{equation}\label{u21}
g_{ij,k}\,X^k+g_{jk}\,X^k_{\,\,,i}+g_{ik}\,X^{k}_{\,\,,j}\,=\,0,
\end{equation}
where $g_{ij}$ denotes the metric tensor components, $X^k$ are the
components of the Killing vector field and the commas in the
subscript are used for partial derivatives with respect to spacetime
coordinates. The homothetic symmetries are obtained by replacing the
right-hand side of equation (\ref{u21}) by $2\,\alpha\,g_{ij}$,
$\alpha$ is an arbitrary constant. Ricci collineation (RC) of a
spacetime is achieved by replacing the metric tensor $g_{ij}$ by the
Ricci tensor $R_{ij}$. This type of collineation satisfies the
equation  \cite{katzin1}:
\begin{equation}\label{u22}
R_{ij,k}\,X^k+R_{jk}\,X^k_{\,\,,i}+R_{ik}\,X^{k}_{\,\,,j}\,=\,0.
\end{equation}
If we take $2\,\alpha\,R_{ij}$ instead of zero on the right-hand
side of equation (\ref{u22}), we get
\begin{equation}\label{u23}
R_{ij,k}\,X^k+R_{jk}\,X^k_{\,\,,i}+R_{ik}\,X^{k}_{\,\,,j}\,=\,2\,\alpha\,R_{ij}.
\end{equation}
If $\alpha$ is an arbitrary constant in equation (\ref{u23}), the
vector field $X$ is known as \textit{homothetic RC} or the
\textit{Ricci inheritance symmetries} \cite{duggal1, duggal2} and if
$\alpha$ is an arbitrary function of the spacetime coordinates, then
$X$ becomes the \textit{conformal Ricci Collineations} \cite{tsam1}.
For Riemannian manifolds, CRC were called concircular vector fields.

The static spherical symmetric spacetimes according to the RC are
classified and the relation between isometries and the RC was
established in \cite{amir1}. The RC  for Bianchi type I, III and
Kantowski-Sachs spacetimes have been found in \cite{camci1}. The RC
of Bainchi type II, VIII and IX spacetimes are presented by Yavuz
and Camci \cite{yavuz1}. A complete classification of cylindrically
symmetric static Lorentzian manifold according to their RC is
provided by Qadir et al \cite{qadir1}. They are also compared with
Killing and homothetic symmetries. Camci and Barnes \cite {camci2}
obtained the RC and Ricci inheritance collineation in the
non-degenrate case for FRW spacetimes. The relation between RC and
isometries for static plane symmetric spacetimes is established in
\cite{farid1}. RC for maximally symmetric transverse spacetimes are
considered for both degenerate and non-degenerate cases in
\cite{akbar1, akbar2}. The Ricci inheritance symmetries of
spherically symmetric spacetimes and Friedman models are discussed
in \cite{bokhari1} and \cite{bokhari2}, respectively.

The RC have been investigated for many spacetimes, however Ricci
inheritance symmetries have been discussed for only few spacetimes.
Recently, Hussain et al \cite{hussain1} and Ali and Khan \cite{ali0} explored
the Ricci inheritance symmetries in Bainchi type I spacetimes and
plane symmetric static spacetimes, respectively. Conformal Killing
vector fields (CKVFs) are symmetries of the metric tensor, while conformal
Ricci collineations (CRC) are symmetries of Ricci tensor. In literature a
relationship between the conformal factors of CKVFs and CRCs is established \cite{ugur1}.
It is also well known that when the Ricci tensor is non-degenerate, the maximum dimension
of the group of CRCs is 15 and this dimension is achieved only if the Ricci tensor
when taken as a metric is conformally flat. Recently, a relationship between CRCs and CKVFs
for pp-waves has been also discussed in detail \cite{Keane1}. Ali and Suhail \cite{ali-0}
solved CKVFs equations along with the equation (\ref{u23}) but they replaced the Ricci tensor $R_{ij}$
in the right hand side by the metric tensor $g_{ij}$. They solved it for static plane-symmetric four dimensional Lorenzian manifold and the vector field is called the concircular vector fields. Our aim of this work is to classify a plane-symmetric static spacetime according to
the conformal Ricci symmetries with the help of the conformal Ricci
collineation equation (\ref{u23}) by taking the conformal factor
$\alpha$ to be a general function of the spacetime coordinates $t$,
$x$, $y$ and $z$. This article is organized as follows: In section
2, ten coupled conformal Ricci collineation equations are obtained.
These equations are solved for non-degenerate and generate cases in
sections 3 and 4, respectively and the conformal Ricci collineations
are obtained. A brief summary of the work and some discussion on the
obtained results are introduced in the last section.

%%%%%%%%%%%%%%%%%%%%%%%%%%%%%%%%%%%%%%%%%%%%%%%
\section{Conformal Ricci collineation equations}
%%%%%%%%%%%%%%%%%%%%%%%%%%%%%%%%%%%%%%%%%%%%%%%

We take the plane-symmetric static spacetime in the the convention
coordinates $(x^0=t,\,x^1=x,\,x^2=y,\,x^3=z)$, in the form
\begin{equation}  \label{u31}
ds^2=-e^{2\,\mu(x)}\,dt^2+dx^2+e^{2\,\nu(x)}\,\left(dy^2+dz^2\right),
\end{equation}
where $\mu$ and $\nu$ are functions of $x$ only. The non-zero Ricci
tensor components for the space time (\ref{u31}) are
\begin{equation}\label{u32}
\left\{
\begin{array}{ll}
R_{00}\,=\,-e^{2\,\mu}\,\left(\mu''+2\,\mu'\,\nu'+\mu'^{2}\right)=A(x),\\
\\
R_{11}\,=\,2\,\left(\nu''+\nu'^{2}\right)-\left(\mu''+\mu'^{2}\right)=B(x),\\
\\
R_{22}\,=\,R_{33}\,=\,e^{2\,\nu}\,\left(\nu''+\mu'\,\nu'+2\,\nu'^{2}\right)=C(x),
\end{array}
\right.
\end{equation}
where the prime denotes the derivative with respect to $x$. Using
Ricci components above in equation (\ref{u23}) yields the following
ten partial differential equations:
\begin{equation}  \label{u33-1}
A'\,X^{1}+2\,A\,X^0_{\,,0}\,=\,2\,\alpha\,A,
\end{equation}
\begin{equation}  \label{u33-2}
A\,X^{0}_{\,,1}+B\,X^1_{\,,0}\,=\,0,
\end{equation}
\begin{equation}  \label{u33-3}
A\,X^{0}_{\,,2}+C\,X^2_{\,,0}\,=\,0,
\end{equation}
\begin{equation}  \label{u33-4}
A\,X^{0}_{\,,3}+C\,X^3_{\,,0}\,=\,0,
\end{equation}
\begin{equation}  \label{u33-5}
B'\,X^{1}+2\,B\,X^1_{\,,1}\,=\,2\,\alpha\,B,
\end{equation}
\begin{equation}  \label{u33-6}
B\,X^{1}_{\,,2}+C\,X^2_{\,,1}\,=\,0,
\end{equation}
\begin{equation}  \label{u33-7}
B\,X^{1}_{\,,3}+C\,X^3_{\,,1}\,=\,0,
\end{equation}
\begin{equation}  \label{u33-8}
C'\,X^{1}+2\,C\,X^2_{\,,2}\,=\,2\,\alpha\,C,
\end{equation}
\begin{equation}  \label{u33-9}
X^{2}_{\,,3}+X^3_{\,,2}\,=\,0,
\end{equation}
\begin{equation}  \label{u33-10}
C'\,X^{1}+2\,C\,X^3_{\,,3}\,=\,2\,\alpha\,C.
\end{equation}

%%%%%%%%%%%%%%%%%%%%%%%%%%%%%%%%%%%%%%%%%%%%%%%%%%%%%%%%
\section{CRCs for non-degenrate Ricci tensor}
%%%%%%%%%%%%%%%%%%%%%%%%%%%%%%%%%%%%%%%%%%%%%%%%%%%%%%%

In this section, we solve Eqs. (\ref{u33-1})--(\ref{u33-10}) by
considering all the Ricci tensor components to be non-zero, that is
$\mathrm{det}\left(R_{ij}\right)\,=\,A(x)\,B(x)\,C^2(x)\,\neq\,0$.
We shall use direct integration technique to find the components
$X^0$, $X^1$, $X^2$ and $X^3$ of the conformal Ricci symmetries and
the function $\alpha$. The process of obtaining these components is
explained as the following:

We differentiate the Eq. (\ref{u33-3}) with respect to $z$, Eq.
(\ref{u33-4}) with respect to $y$ and Eq. (\ref{u33-9}) with respect
to $t$ to obtain a relation
\begin{equation}  \label{u41-1}
X^0_{\,,yz}\,=\,X^2_{\,,tz}\,=\,X^3_{\,,ty}\,=\,0.
\end{equation}
Similarly, Eqs. (\ref{u33-6}), (\ref{u33-7}) and (\ref{u33-9}),
after differentiating with respect to $z$, $y$, $x$, respectively
and some simple algebraic calculation, gives
\begin{equation}  \label{u41-2}
X^1_{\,,yz}\,=\,X^2_{\,,xz}\,=\,X^3_{\,,xy}\,=\,0.
\end{equation}
By applicable a similar manor between Eqs. (\ref{u33-3}),
(\ref{u33-4}), (\ref{u33-6}), (\ref{u33-7}), (\ref{u33-8}),
(\ref{u33-9}) and (\ref{u33-10}), gets
\begin{equation}  \label{u41-3}
X^i_{\,,yy}\,-\,X^i_{\,,zz}\,=\,X^j_{\,,yy}\,+\,X^j_{\,,zz}\,=\,0,\,\,\,i=0,1,\,\,\,j=2,3.
\end{equation}
Integrating Eqs. (\ref{u41-1}) and (\ref{u41-2}), substituting the
results in (\ref{u41-3}), again integrate (\ref{u41-3}) and solving
the Eqs. (\ref{u33-1}), (\ref{u33-3}), (\ref{u33-4}), (\ref{u33-6}),
(\ref{u33-7}) and (\ref{u33-9}), we get
\begin{equation}  \label{u41-4}
\left\{
  \begin{array}{ll}
    X^0\,=\,F^0-2\,\sqrt{\dfrac{C(x)}{A(x)}}\,\Big[y\,G^1+z\,G^2+\left(y^2+z^2\right)\,G^3\Big]_{,t},\\
\\
  X^1\,=\,F^1-\dfrac{C(x)}{B(x)}\,\Bigg(\left[\dfrac{A(x)}{C(x)}\right]'\,\sqrt{\dfrac{C(x)}{A(x)}}\,\Big[y\,G^1+z\,G^2+\left(y^2+z^2\right)\,G^3\Big]\\
\,\,\,\,\,\,\,\,\,\,\,\,\,\,\,\,\,\,\,\,\,\,\,\,\,\,\,\,\,\,\,\,\,\,\,\,\,\,\,\,\,\,\,\,\,\,\,\,\,\,\,\,\,\,\,\,\,\,\,\,\,\,
\,\,\,\,\,\,\,\,\,\,\,\,\,\,\,\,\,\,\,\,\,\,\,\,\,\,\,\,\,\,\,\,\,
+\Big[y\,H^1+z\,H^2+\left(y^2+z^2\right)\,H^3\Big]_{,x}\Bigg),\\
\\
  X^2\,=\,H^1+2\,H^3\,y+d_0\,z+d_1\,\left(y^2-z^2\right)+2\,d_2\,y\,z+2\,\sqrt{\dfrac{A(x)}{C(x)}}\,\Big[G^1+2\,G^3\,y\Big],\\
\\
X^3\,=\,H^2-d_0\,y+2\,H^3\,z+d_2\,\left(z^2-y^2\right)+2\,d_1\,y\,z+2\,\sqrt{\dfrac{A(x)}{C(x)}}\,\Big[G^2+2\,G^3\,z\Big],
  \end{array}
\right.
\end{equation}
and
\begin{equation}  \label{u41-4-0}
\alpha\,=\,X^0_{,t}+\dfrac{A'(x)\,X^1}{2\,A(x)},
\end{equation}
where $F^i=F^i(t,x)$, $G^j=G^j(t)$ and $H^j=H^j(x)$ are functions of
integration while $d_k$, are constants of integration for all
$i=0,1$, $j=1,2,3$ and $k=0,1,2$. Substituting these values of
$X^0$, $X^1$, $X^2$ and $X^3$ in the system
(\ref{u33-1})-(\ref{u33-10}), expanding it with the aid of
\textit{Mathematica Program} and set the coefficients involving $y$
and $z$ and various products equal zero, give to the following set
of over-determined equations:
\begin{equation}  \label{u410-1}
A\,F^0_{,x}+B\,F^1_{,t}\,=\,0,
\end{equation}
\begin{equation}  \label{u410-2}
\left[\dfrac{B}{A}\right]'\,F^1+2\,\left[\dfrac{B}{A}\right]\,\Big(F^1_{,x}-F^0_{,t}\Big)\,=\,0,
\end{equation}
\begin{equation}  \label{u410-3}
\left[\dfrac{A}{C}\right]'\,F^1-2\,\left[\dfrac{A}{C}\right]\left[4\,G^3\,\sqrt{\dfrac{A}{C}}+2\,H^3-F^0_{,t}\right]\,=\,0,
\end{equation}
\begin{equation}  \label{u410-4}
%\left\{
  \begin{array}{ll}
    \left(\dfrac{A^2}{B\,C}\left[\dfrac{C}{A}\right]^{\prime\,2}\right)'\,G^j+
\left[\dfrac{C}{A}\right]'\,
\Bigg(\left[\dfrac{A}{C}\right]^{3/2}\,\left[
\dfrac{C^2}{A\,B}\right]'\,H^j_{,x}\\
\\
\,\,\,\,\,\,\,\,\,\,\,\,\,\,\,\,\,\,\,\,\,\,\,\,\,\,\,\,\,\,\,\,\,\,\,\,\,\,\,\,
\,\,\,\,\,\,\,\,\,\,\,\,\,\,\,\,\,\,\,\,\,\,\,\,\,\,\,\,\,\,\,\,\,\,\,\,\,\,\,\,
-2\,\left[\dfrac{\sqrt{A\,C}}{B}\right]\,H^j_{,xx}-4\,G^j_{,tt}\Bigg)\,=\,0,\,\,\,j=1,2,3,
  \end{array}
%\right.
\end{equation}
\begin{equation}  \label{u410-5}
\left[\dfrac{A}{C}\right]^{\prime\,2}\,G^j+\left[\dfrac{A}{C}\right]'\,\sqrt{\dfrac{A}{C}}\,H^j_{,x}+\left[\dfrac{4\,A\,B}{C^2}\right]\,\left[
d_j\,\sqrt{\dfrac{A}{C}}+G^j_{,tt}\right]\,=\,0,\,\,\,j=1,2,3,
\end{equation}
where $d_3\,=\,0$. Differentiating the Eq. (\ref{u410-5}) with
respect to $t$, the following equation is obtained:
\begin{equation}  \label{u412}
\left[\dfrac{A}{C}\right]^{\prime\,2}\,G^j_{,t}+\left[\dfrac{4\,A\,B}{C^2}\right]\,\left[
G^j_{,ttt}\right]\,=\,0,\,\,\,j=1,2,3.
\end{equation}
The above equation gives rise to the following two possibilities:
(I): $\left[\dfrac{A(x)}{C(x)}\right]'\,\neq\,0$; (II):
$\left[\dfrac{A(x)}{C(x)}\right]'\,=\,0$. We shall discuss each case
in turn.

\textbf{Case (I):} In this case we consider
$\left[\dfrac{A(x)}{C(x)}\right]'\,\neq\,0$. Then Eq. (\ref{u412})
leads to
\begin{equation}  \label{u413}
B(x)=\dfrac{C^2(x)}{4\,c_0\,A(x)}\,\left[\dfrac{A(x)}{C(x)}\right]^{\prime\,2},
\end{equation}
where $c_0$ is an arbitrary constant. Now we study two subcases: The
first one when $c_0$ positive and the second when $c_0$ negative as
follows:

\textbf{Case (I-A):} Here, taking $c_0\,=\,a_0^2\,>\,0$ and solving
Eq. (\ref{u412}) we get:
\begin{equation}  \label{u414}
G^j(t)=e_{j1}+e_{j2}\,\cos\left[a_0\,t\right]+e_{j3}\,\sin\left[a_0\,t\right],\,\,\,\,\,\,\,j=1,2,3,
\end{equation}
where $e_{ij}$ are constants of integration for all $i,j\,=\,1,2,3$.
Substituting the values of $B(x)$ and $G^j(t)$ in Eq. (\ref{u410-4})
and solving the resulting equation we obtain:
\begin{equation}  \label{u415}
H^j(x)\,=\,f_{j}-\dfrac{d_j\,A(x)}{a_0^2\,C(x)}-2\,e_{j1}\,\sqrt{\dfrac{A(x)}{B(x)}},\,\,\,\,\,\,\,j=1,2,3,
\end{equation}
where $f_j$ are constants of integration for all $j=1,2,3$. From Eq.
(\ref{u410-3}) we have
\begin{equation}  \label{u416}
F^1(t,x)=\dfrac{A(x)\,C(x)\left(4\,f_3-2F^0_{,t}\right)-8\,A(x)\sqrt{A(x)\,C(x)}
\,\left(e_{32}\,\cos\left[a_0\,t\right]+e_{33}\,\sin\left[a_0\,t\right]\right)}{A(x)\,C'(x)-C(x)\,A'(x)}.
\end{equation}
and from Eq. (\ref{u410-2}) we get
\begin{equation}  \label{u417}
F^0(t,x)=H^4(x)+G^4(t)\,\sqrt{\dfrac{C(x)}{A(x)}}+2\,\sqrt{\dfrac{A(x)}{C(x)}}\Big[
\left(\dfrac{e_{32}}{a_0}\right)\,\sin\left[a_0\,t\right]-\left(\dfrac{e_{33}}{a_0}\right)\,\cos\left[a_0\,t\right]\Big],
\end{equation}
where $G^4(t)$ and $H^4(x)$ are functions of integration. Now, Eq.
(\ref{u410-1}) becomes:
\begin{equation}  \label{u418}
\left[\dfrac{A(x)}{C(x)}\right]'\left[G^4_{,tt}(t)+a_0^2\,G^4(t)\right]\,=\,2\,a_0^2\,\left[\dfrac{A(x)}{C(x)}\right]^{3/2}\,H^4_{,x}.
\end{equation}
Solving the above equation we obtain
\begin{equation}  \label{u419}
G^4(t)=e_{41}+e_{42}\,\cos\left[a_0\,t\right]+e_{43}\,\sin\left[a_0\,t\right],
\end{equation}
\begin{equation}  \label{u420}
H^4(x)=f_{4}-e_{41}\,\sqrt{\dfrac{C(x)}{A(x)}},
\end{equation}
where $f_4$ and $e_{4j}$ are constants of integration for all
$j\,=\,1,2,3$. Hence, we have the following components of CRCs:
\begin{equation}  \label{u41-CRC-IA}
\left\{
  \begin{array}{ll}
    X^0\,=\,a_1+\sqrt{\dfrac{C(x)}{A(x)}}\,\Bigg[\left(
a_2+a_3\,y+a_4\,z+a_5\,\left[a_0^2\,(y^2+z^2)+\dfrac{A(x)}{C(x)}\right]\right)\,\cos\left[a_0\,t\right]\\
\\
\,\,\,\,\,\,\,\,\,\,\,\,\,\,\,\,\,\,\,\,\,\,\,\,\,\,\,\,\,\,\,\,\,\,\,\,\,\,\,\,\,\,\,\,\,
+\left(
a_6+a_7\,y+a_8\,z+a_9\,\left[a_0^2\,(y^2+z^2)+\dfrac{A(x)}{C(x)}\right]\right)\,\sin\left[a_0\,t\right]\Bigg],\\
\\
X^1\,=\,\dfrac{2\,a_0\,C(x)\,\sqrt{C(x)}}{\sqrt{A(x)}\,\left[A(x)\,C'(x)-C(x)\,A'(x)\right]}\,\Bigg[a_{10}+a_{11}\,y+a_{12}\,z\\
\\
\,\,\,\,\,\,\,\,\,\,\,\,\,\,\,\,\,\,\,\,\,\,\,\,\,\,\,\,\,\,\,\,\,\,\,\,\,\,\,\,
+\left(
a_6+a_7\,y+a_8\,z+a_9\,\left[a_0^2\,(y^2+z^2)-\dfrac{A(x)}{C(x)}\right]\right)\,\cos\left[a_0\,t\right]\\
\\
\,\,\,\,\,\,\,\,\,\,\,\,\,\,\,\,\,\,\,\,\,\,\,\,\,\,\,\,\,\,\,\,\,\,\,\,\,\,\,\,
-\left(
a_2+a_3\,y+a_4\,z+a_5\,\left[a_0^2\,(y^2+z^2)-\dfrac{A(x)}{C(x)}\right]\right)\,\sin\left[a_0\,t\right]\Bigg],\\
\\
X^2=\dfrac{a_0}{2}\,\left[a_{13}-2\,a_{10}\,y+a_{14}\,z+a_{11}\left(z^2-y^2+\dfrac{A(x)}{a_0^2\,C(x)}\right)\right]\\
\\
\,\,\,\,\,\,\,\,\,\,\,\,\,\,\,\,\,\,\,\,
+\sqrt{\dfrac{A(x)}{C(x)}}\,\Bigg[\left(\dfrac{h_7}{a_0}+2\,a_0\,a_9\,y\right)\,\cos\left[a_0\,t\right]-\left(
\dfrac{a_3}{a_0}+2\,a_0\,a_5\,y\right)\,\sin\left[a_0\,t\right]\Bigg],\\
\\
X^3=\dfrac{a_0}{2}\,\left[a_{15}-a_{14}\,y-2\,a_{10}\,z+a_{12}\left(y^2-z^2+\dfrac{A(x)}{a_0^2\,C(x)}\right)\right]\\
\\
\,\,\,\,\,\,\,\,\,\,\,\,\,\,\,\,\,\,\,\,
+\sqrt{\dfrac{A(x)}{C(x)}}\,\Bigg[\left(\dfrac{h_8}{a_0}+2\,a_0\,a_9\,z\right)\,\cos\left[a_0\,t\right]-\left(
\dfrac{h_4}{a_0}+2\,a_0\,a_5\,z\right)\,\sin\left[a_0\,t\right]\Bigg],
  \end{array}
\right.
\end{equation}
and
\begin{equation}  \label{u41-CRC-IA-0}
%\left\{
  \begin{array}{ll}
\alpha\,=\,\dfrac{a_0\,\sqrt{A(x)\,C(x)}}{A(x)\,C'(x)-C(x)\,A'(x)}\,\Bigg[
\left(a_{10}+a_{11}\,y+a_{12}\,z\right)\,\left(\dfrac{\sqrt{C(x)}\,A'(x)}{\sqrt{A(x)}}\right)\\
\\
\,\,\,\,\,\,\,\,\,\,\,\,\,\,\,\,\,\,\,\,\,\,\,\,\,\,\,\,\,+\Bigg(
\left[a_6+a_7\,y+a_8\,z\right]\,C'(x)+a_9\,\Big[a_0^2\,(y^2+z^2)\,C'(x)\\
\\
\,\,\,\,\,\,\,\,\,\,\,\,\,\,\,\,\,\,\,\,\,\,\,\,\,\,\,\,\,-2\,A'(x)+\dfrac{A(x)\,C'(x)}{C(x)}\Big]\Bigg)\,\cos\left[a_0\,t\right]
-\Bigg(
\left[a_2+a_3\,y+a_4\,z\right]\,C'(x)\\
\\
\,\,\,\,\,\,\,\,\,\,\,\,\,\,\,\,\,\,\,\,\,\,\,\,\,\,\,\,\,+a_5\,\left[a_0^2\,(y^2+z^2)\,C'(x)-2\,A'(x)+\dfrac{A(x)\,C'(x)}{C(x)}\right]\Bigg)\,\sin\left[a_0\,t\right]\Bigg],
  \end{array}
%\right.
\end{equation}
where
$B(x)=\dfrac{A(x)}{4\,a_0^2}\,\left(\dfrac{A'(x)}{A(x)}-\dfrac{C'(x)}{C(x)}\right)^2$
and $a_i$, $i=0,1,...,15$ are arbitrary constants of integration.

\textbf{Case (I-B):} In this case we take $c_0\,=\,-a_0^2\,<\,0$. By
a similar method, we obtain the following components of CRCs:
\begin{equation}  \label{u41-CRC-IB}
\left\{
  \begin{array}{ll}
    X^0\,=\,a_1+\sqrt{\dfrac{C(x)}{A(x)}}\,\Bigg[\left(
a_2+a_3\,y+a_4\,z-a_5\,\left[a_0^2\,(y^2+z^2)-\dfrac{A(x)}{C(x)}\right]\right)\,\cosh\left[a_0\,t\right]\\
\\
\,\,\,\,\,\,\,\,\,\,\,\,\,\,\,\,\,\,\,\,\,\,\,\,\,\,\,\,\,\,\,\,\,\,\,\,\,\,\,\,\,\,\,\,\,
+\left(
a_6+a_7\,y+a_8\,z-a_9\,\left[a_0^2\,(y^2+z^2)-\dfrac{A(x)}{C(x)}\right]\right)\,\sinh\left[a_0\,t\right]\Bigg],\\
\\
X^1\,=\,\dfrac{2\,a_0\,C(x)\,\sqrt{C(x)}}{\sqrt{A(x)}\,\left[A(x)\,C'(x)-C(x)\,A'(x)\right]}\,\Bigg[a_{10}+a_{11}\,y+a_{12}\,z\\
\\
\,\,\,\,\,\,\,\,\,\,\,\,\,\,\,\,\,\,\,\,\,\,\,\,\,\,\,\,\,\,\,\,\,\,\,\,\,\,\,\,
+\left(
a_6+a_7\,y+a_8\,z-a_9\,\left[a_0^2\,(y^2+z^2)+\dfrac{A(x)}{C(x)}\right]\right)\,\cosh\left[a_0\,t\right]\\
\\
\,\,\,\,\,\,\,\,\,\,\,\,\,\,\,\,\,\,\,\,\,\,\,\,\,\,\,\,\,\,\,\,\,\,\,\,\,\,\,\,
-\left(
a_2+a_3\,y+a_4\,z-a_5\,\left[a_0^2\,(y^2+z^2)+\dfrac{A(x)}{C(x)}\right]\right)\,\sinh\left[a_0\,t\right]\Bigg],\\
\\
X^2=\dfrac{a_0}{2}\,\left[a_{13}-2\,a_{10}\,y+a_{14}\,z+a_{11}\left(z^2-y^2-\dfrac{A(x)}{a_0^2\,C(x)}\right)\right]\\
\\
\,\,\,\,\,\,\,\,\,\,\,\,\,\,\,\,\,\,\,\,
-\sqrt{\dfrac{A(x)}{C(x)}}\,\Bigg[\left(\dfrac{h_7}{a_0}-2\,a_0\,a_9\,y\right)\,\cosh\left[a_0\,t\right]-\left(
\dfrac{a_3}{a_0}-2\,a_0\,a_5\,y\right)\,\sinh\left[a_0\,t\right]\Bigg],\\
\\
X^3=\dfrac{a_0}{2}\,\left[a_{15}-a_{14}\,y-2\,a_{10}\,z+a_{12}\left(y^2-z^2-\dfrac{A(x)}{a_0^2\,C(x)}\right)\right]\\
\\
\,\,\,\,\,\,\,\,\,\,\,\,\,\,\,\,\,\,\,\,
-\sqrt{\dfrac{A(x)}{C(x)}}\,\Bigg[\left(\dfrac{h_8}{a_0}-2\,a_0\,a_9\,z\right)\,\cosh\left[a_0\,t\right]-\left(
\dfrac{h_4}{a_0}-2\,a_0\,a_5\,z\right)\,\sinh\left[a_0\,t\right]\Bigg],
\end{array}
\right.
\end{equation}
and
\begin{equation}  \label{u41-CRC-IB-0}
%\left\{
  \begin{array}{ll}
\alpha\,=\,\dfrac{a_0\,\sqrt{A(x)\,C(x)}}{A(x)\,C'(x)-C(x)\,A'(x)}\,\Bigg[
\left(a_{10}+a_{11}\,y+a_{12}\,z\right)\,\left(\dfrac{\sqrt{C(x)}\,A'(x)}{\sqrt{A(x)}}\right)\\
\\
\,\,\,\,\,\,\,\,\,\,\,\,\,\,\,\,\,\,\,\,\,\,\,
+\Bigg(
\left[a_6+a_7\,y+a_8\,z\right]\,C'(x)-a_9\,\Big[a_0^2\,(y^2+z^2)\,C'(x)\\
\\
\,\,\,\,\,\,\,\,\,\,\,\,\,\,\,\,\,\,\,\,\,\,\,
+2\,A'(x)+\dfrac{A(x)\,C'(x)}{C(x)}\Big]\Bigg)\,\cosh\left[a_0\,t\right]-\Bigg(
\left[a_2+a_3\,y+a_4\,z\right]\,C'(x)\\
\\
\,\,\,\,\,\,\,\,\,\,\,\,\,\,\,\,\,\,\,\,\,\,\,
-a_5\,\left[a_0^2\,(y^2+z^2)\,C'(x)+2\,A'(x)+\dfrac{A(x)\,C'(x)}{C(x)}\right]\Bigg)\,\sinh\left[a_0\,t\right]\Bigg],
  \end{array}
%\right.
\end{equation}
where
$B(x)=-\dfrac{A(x)}{4\,a_0^2}\,\left(\dfrac{A'(x)}{A(x)}-\dfrac{C'(x)}{C(x)}\right)^2$
and $a_i$, $i=0,1,...,15$ are arbitrary constants of integration.

\textbf{Case (II):} In this case we consider
$\left[\dfrac{A(x)}{C(x)}\right]'\,=\,0$. Then the components of
CRCs are given by:

\begin{equation}  \label{u41-CRC-II}
\left\{
  \begin{array}{ll}
    X^0\,=\,a_1+a_2\,t+a_3\,y+a_4\,z+a_5\,\left[a_0\,t^2-y^2-z^2-\left(\int\sqrt{\dfrac{B(x)}{C(x)}}\,dx\right)^2\right]\\
\\
\,\,\,\,\,\,\,\,\,\,\,\,\,\,\,\,\,\,\,\,\,\,\,\,\,\,\,\,\,\,\,\,\,\,\,\,\,\,\,\,\,\,\,\,\,\,\,\,\,\,\,\,\,\,\,\,\,\,\,\,\,\,\,\,\,\,\,\,\,\,
+2\,a_6\,t\,y+2\,a_7\,t\,z+\left(a_8+2\,a_9\,t\right)\,\left(\int\sqrt{\dfrac{B(x)}{C(x)}}\,dx\right),\\
\\
X^1\,=\,\sqrt{\dfrac{C(x)}{B(x)}}\,\Bigg[a_{10}-a_0\,a_8\,t+a_{11}\,y+a_{12}\,z\\
\\
\,\,\,\,\,\,\,\,\,\,\,\,\,\,\,\,\,\,\,\,\,\,\,\,\,\,\,\,\,\,\,\,\,\,\,\,\,\,\,\,\,\,\,\,\,\,\,
-a_9\,\left[a_0\,t^2+y^2+z^2-\left(\int\sqrt{\dfrac{B(x)}{C(x)}}\,dx\right)^2\right]\\
\\
\,\,\,\,\,\,\,\,\,\,\,\,\,\,\,\,\,\,\,\,\,\,\,\,\,\,\,\,\,\,\,\,\,\,\,\,\,\,\,\,\,\,\,\,\,\,\,
+\left(a_2\,+2\,a_0\,a_5\,t+2\,a_6\,y+2\,a_7\,z\right)\,\left(\int\sqrt{\dfrac{B(x)}{C(x)}}\,dx\right)\Bigg],\\
\\
X^2=\,a_{13}-a_0\,a_3\,t+a_2\,y+a_{14}\,z-a_6\,\left[a_0\,t^2+y^2+z^2-\left(\int\sqrt{\dfrac{B(x)}{C(x)}}\,dx\right)^2\right]\\
\\
\,\,\,\,\,\,\,\,\,\,\,\,\,\,\,\,\,\,\,\,\,\,\,\,\,\,\,\,\,\,\,\,\,\,\,\,\,\,\,\,\,\,\,\,\,\,\,\,\,\,\,\,\,\,\,\,\,\,\,\,
+2\,a_0\,a_5\,t\,y+2\,a_7\,y\,z-\left(a_{11}-2\,a_9\,y\right)\,\left(\int\sqrt{\dfrac{B(x)}{C(x)}}\,dx\right),\\
\\
X^3=\,a_{15}-a_0\,a_4\,t-a_{14}\,y+a_{2}\,z-a_7\,\left[a_0\,t^2+y^2+z^2+\left(\int\sqrt{\dfrac{B(x)}{C(x)}}\,dx\right)^2\right]\\
\\
\,\,\,\,\,\,\,\,\,\,\,\,\,\,\,\,\,\,\,\,\,\,\,\,\,\,\,\,\,\,\,\,\,\,\,\,\,\,\,\,\,\,\,\,\,\,\,\,\,\,\,\,\,\,\,\,\,\,\,\,
+2\,a_0\,a_5\,t\,z+2\,a_6\,y\,z-\left(a_{12}-2\,a_9\,z\right)\,\left(\int\sqrt{\dfrac{B(x)}{C(x)}}\,dx\right),
  \end{array}
\right.
\end{equation}
and
\begin{equation}  \label{u41-CRC-II1-0}
%\left\{
  \begin{array}{ll}
    \alpha\,=\,\left(a_2+2\,a_0\,a_5\,t+2\,a_6\,y+2\,a_7\,z\right)\,\left[1+\left(\dfrac{C'(x)}{2\,\sqrt{B(x)\,C(x)}}\right)\int\sqrt{\dfrac{B(x)}{C(x)}}\,dx\right]\\
\\
\,\,\,\,\,\,\,\,\,\,\,\,\,\,\,\,\,\,\,\,\,\,\,\,\,\,\,\,\,\,\,\,\,\,\,
+\dfrac{\left(a_{10}-a_0\,a_8\,\,t+a_{11}\,y+a_{12}\,z\right)\,C'(x)}{2\,\sqrt{B(x)\,C(x)}},\\
\\
\,\,\,
+h_9\left(2\,\int\sqrt{\dfrac{B(x)}{C(x)}}\,dx-\dfrac{C'(x)}{2\,\sqrt{B(x)\,C(x)}}\,\left[a_0\,t^2+y^2+z^2-\left(\int\sqrt{\dfrac{B(x)}{C(x)}}\,dx\right)^2\right]\right),
  \end{array}
%\right.
\end{equation}
where $A(x)=a_0\,C(x)$ and $a_i$, $i=0,1,...,15$ are arbitrary
constants.

%%%%%%%%%%%%%%%%%%%%%%%%%%%%%%%%%%%%%%%%%%%%%%%%%%%%%%%%
\section{CRCs for degenrate Ricci tensor}
%%%%%%%%%%%%%%%%%%%%%%%%%%%%%%%%%%%%%%%%%%%%%%%%%%%%%%%

In this case, the Ricci tensor $R_{ij}$ is degenerate, that is,
$\mathrm{det}(R_{ij})\,=\,0$, i.e., $A(x)\,B(x)\,C(x)\,=\,0$, then
there are six possible cases depending on whether one or two
components of the Ricci tensor are zero. The conformal Ricci
collineations are given in each of these cases; however we omit the
basic calculation and give the final form of CRCs in each case.

\textbf{Case (IV):} In this case $A(x)\,=B(x)\,=\,0$ and
$C(x)\,\neq\,0$. We are left with the following seven equations:
\begin{equation}  \label{u51-1}
C'\,X^{1}+2\,C\,X^2_y\,=\,2\,\alpha\,C,
\end{equation}
\begin{equation}  \label{u51-2}
C'\,X^{1}+2\,C\,X^3_z\,=\,2\,\alpha\,C,
\end{equation}
\begin{equation}  \label{u51-3}
X^2_t\,=\,X^2_x\,=\,X^3_t\,=\,X^3_x\,=\,0,
\end{equation}
\begin{equation}  \label{u51-4}
X^2_z+X^3_y\,=\,0.
\end{equation}
Eqs. (\ref{u51-3}) and (\ref{u51-4}) give $X^2=F_y(y,z)$ and
$X^3=G(z)-F_{z}(y,x)$. Subtracting Eqs. (\ref{u51-1}) and
(\ref{u51-2}) and differentiating the resulting equation with
respect to $y$, we get an equation $F_{yyy}+F_{yzz}\,=\,0$. Now, we
have the general solution of conformal Ricci collineations as the
following:
\begin{equation}  \label{u51-CRC-IV}
\left\{
  \begin{array}{ll}
    X^0\,=\,X^0(t,x,y,z),\,\,\,\,\,\,\,\,\,\,X^1\,=\,X^1(t,x,y,z),\\
\\
X^2\,=\,\Psi_{,y}(y,z),\,\,\,\,\,\,\,\,\,\,\,\,\,\,\,\,\,\,\,\,\,X^3\,=\,-\Psi_{,z}(y,z),
  \end{array}
\right.
\end{equation}
and $\alpha=\Psi_{,yy}+\dfrac{C'(x)\,X^1}{2\,C(x)}$, where
$X^0(t,x,y,z)$, $X^1(t,x,y,z)$ and $C(x)$ are arbitrary functions
while $\Psi(y,z)$ is a function satisfy the relation
$\Psi_{,yy}+\Psi_{,zz}=0$.

\textbf{Case (V):} In this case we take $A(x)\,=C(x)\,=\,0$ and
$B(x)\,\neq\,0$. Here we are left with the following four equations:
\begin{equation}  \label{u52-1}
B'\,X^{1}+2\,C\,X^1_x\,=\,2\,\alpha\,B,
\end{equation}
\begin{equation}  \label{u52-2}
X^{1}_{,i}\,=\,0,\,\,\,\mathrm{where}\,\,\,i=0,2,3.
\end{equation}
Solving these four equations we observe the three (one temporal and
two spatial) conformal Ricci collineations are arbitrary functions
of the spacetime coordinates $(t,x,y,z)$ and the spatial CRC $X^1$
is given as
\begin{equation}  \label{u51-CRC-V}
%\left\{
  \begin{array}{ll}
    X^1\,=\,\dfrac{a_1+\int\,\alpha(x)\,\sqrt{B(x)}\,dx}{\sqrt{B(x)}},
  \end{array}
%\right.
\end{equation}
and the conformal factor $\alpha=\alpha(x)$ is a function of $x$
only, where $a_1$ is an arbitrary constant. In this case the
differential constraints are solved completely and the metric which
admit the above CRC is also obtained as follows:
\begin{equation}  \label{u31-MIIIB}
ds^2\,=\,-a_0^2\,\left(x_0+x\right)^{2\,\mu_0}\,dt^2+dx^2+c_0^2\,\left(x_0+x\right)^{1-\,\mu_0}\,\left(dy^2+dz^2\right),
\end{equation}
where $a_0$, $c_0$, $x_0$ and $\mu_0$ are arbitrary constants. Therefore, the Ricci tensor form can be written as:\\
\begin{equation}  \label{u31-RIIIB}
ds_{Ric}^2\,=\,A(x)\,dt^2+\,B(x)\,dx^2+C(x)\,\left(dy^2+dz^2\right)\,
=\,\left[\dfrac{\left(1-\mu_0\right)\,\left(1+3\,\mu_0\right)}{2\left(x+x_0\right)^2}\right]\,dx^2.
\end{equation}
The above Ricci metric admits conformal Ricci vector field in the
form
\begin{equation}  \label{u31-XIIIB}
X\,=\,X^0(t,x,y,z)\,\dfrac{\partial}{\partial
t}+\left(x+x_0\right)\,\left[\alpha_0+\int\dfrac{\alpha(x)}{x+x_0}\,dx\right]\,\dfrac{\partial}{\partial
x}+\sum_{i=2}^3\,X^i(t,x,y,z)\,\dfrac{\partial}{\partial x^i},
\end{equation}
where $\alpha_0$ is an arbitrary constant while $X^0$, $X^2$ and
$X^3$ are arbitrary functions of the coordinates $(t,x,y,z)$.

\textbf{Case (VI):} In this case we consider $B(x)\,=C(x)\,=\,0$ and
$A(x)\,\neq\,0$. Here we are left with the following equations to be
dealt with:
\begin{equation}  \label{u53-1}
A'\,X^{1}+2\,A\,X^0_t\,=\,2\,\alpha\,A,
\end{equation}
\begin{equation}  \label{u53-2}
X^{0}_{,i}\,=\,0,\,\,\,\mathrm{where}\,\,\,i=1,2,3.
\end{equation}
Solving the above equations we get the following CRCs.
\begin{equation}  \label{u51-CRC-VI}
%\left\{
  \begin{array}{ll}
    X^0\,=\,\Psi(t),\,\,\,\,\,\,\,\,\,\,X^i\,=\,X^i(t,x,y,z),\,\,\mathrm{for}\,\,i=1,2,3,
  \end{array}
%\right.
\end{equation}
and $\alpha=\Psi'(t)+\dfrac{A'(x)\,X^1}{2\,A(x)}$ where $\Psi(t)$,
$X^1(t,x,y,z)$, $X^2(t,x,y,z)$, $X^3(t,x,y,z)$ and $A(x)$ are
arbitrary functions.

\textbf{Case (VII):} In this case $A(x)\,\neq\,0$, $B(x)\,\neq\,0$
and $C(x)\,=\,0$. Then the Eqs. (\ref{u33-1})--(\ref{u33-10})
reduced as follows:
\begin{equation}  \label{u54-1}
A'\,X^{1}+2\,A\,X^0_t\,=\,2\,\alpha\,A,
\end{equation}
\begin{equation}  \label{u54-2}
B'\,X^{1}+2\,B\,X^1_x\,=\,2\,\alpha\,B,
\end{equation}
\begin{equation}  \label{u54-3}
A\,X^0_x+B\,X^1_t\,=\,0,
\end{equation}
\begin{equation}  \label{u54-4}
X^0_y\,=\,X^0_z\,=\,X^1_y\,=\,x^1_z\,=\,0,
\end{equation}
Solving the Eqs. (\ref{u54-1})--(\ref{u54-4}), we get the components
of CRCs as the following:
\begin{equation}  \label{u41-CRC-VII}
\left\{
  \begin{array}{ll}
    X^0\,=\,G(t)-\int\dfrac{B(x)}{A(x)}\,X^1_{,t}(t,x)\,dx,\,\,\,\,\,\,\,\,\,\,\,\,\,\,\,X^1\,=\,X^1(t,x),\\
\\
X^2\,=\,X^2(t,x,y,z),\,\,\,\,\,\,\,\,\,\,\,\,\,\,\,\,\,\,\,\,\,\,\,\,\,\,\,
\,\,\,\,\,\,\,\,\,\,\,\,\,\,\,\,\,\,\,\,\,\,\,\,X^3\,=\,X^3(t,x,y,z),
  \end{array}
\right.
\end{equation}
and $\alpha=X^1_{,x}+\dfrac{B'(x)\,X^1}{2\,B(x)}$ where $G(t)$,
$A(x)$, $B(x)$, $X^2(t,x,y,z)$ and $X^3(t,x,y,z)$ are arbitrary
functions while $X^1(t,x)$ satisfy the following partial integral
differential equation:
\begin{equation}  \label{u41-CRC-VII0}
\begin{array}{ll}
\left(\dfrac{A'(x)}{A(x)}-\dfrac{B'(x)}{B(x)}\right)\,X^1-x^1_{,x}\,=\,G(t)+\int\dfrac{B(x)}{A(x)}\,X^1_{,tt}\,dx.
 \end{array}
\end{equation}

\textbf{Case (VIII):} In this case $A(x)\,\neq\,0$, $C(x)\,\neq\,0$
and $B(x)\,=\,0$. Then the CRC equations are
\begin{equation}  \label{u55-1}
A'\,X^{1}+2\,A\,X^0_t\,=\,2\,\alpha\,A,
\end{equation}
\begin{equation}  \label{u55-2}
C'\,X^{1}+2\,C\,X^2_y\,=\,2\,\alpha\,C,
\end{equation}
\begin{equation}  \label{u55-3}
C'\,X^{1}+2\,C\,X^3_z\,=\,2\,\alpha\,C,
\end{equation}
\begin{equation}  \label{u55-4}
A\,X^0_y+C\,X^2_t\,=\,0,
\end{equation}
\begin{equation}  \label{u55-5}
A\,X^0_z+C\,X^3_t\,=\,0,
\end{equation}
\begin{equation}  \label{u55-6}
X^2_z+X^3_y\,=\,0,
\end{equation}
\begin{equation}  \label{u55-7}
X^i_x\,=\,0,\,\,\,\,\,i=0,2,3.
\end{equation}
We omit to write the basic steps involved in the solution of the
above equations. After some mathematical calculations we obtain the
following two solutions of the components of CRCs:

\textbf{(VIII-1):}
\begin{equation}  \label{u41-CRC-VIII1}
\left\{
  \begin{array}{ll}
    X^0\,=\,\Psi(t)-y\,\Phi(t)-z\,\Omega(t),\,\,\,\,\,\,\,\,\,\,\,\,\,\,\,\,\,\,X^1\,=\,X^1(t,x,y,z),\\
\\
X^2\,=\,a_0\,\Phi(t)+F_{,y}(y,z),\,\,\,\,\,\,\,\,\,\,\,\,\,\,\,\,\,\,\,\,\,\,\,\,\,\,\,\,
X^3\,=\,a_0\,\Omega(t)-F_{,z}(y,z),
  \end{array}
\right.
\end{equation}
and $\alpha=\dfrac{C'\,X^1}{2\,C}+F_{,yy}$ such that $A(x)=a_0\,C(x)$ and the function $F(y,z)$ satisfy $F_{,yy}(y,z)+F_{,zz}(y,z)\,=\,0$ where $C(x)$, $X^1(t,x,y,z)$, $\Phi(t)$ and $\Omega(t)$ are arbitrary functions while $a_0$ is an arbitrary constant.

\textbf{(VIII-2):}
\begin{equation}  \label{u41-CRC-VIII2}
%\left\{
  \begin{array}{ll}
    X^0\,=\,\Psi(t),\,\,\,\,\,\,\,X^1\,=\,X^1(t,x,y,z),\,\,\,\,\,\,\,
    X^2\,=\,F_{,y}(y,z),\,\,\,\,\,\,\,
X^3\,=\,-F_{,z}(y,z),
  \end{array}
%\right.
\end{equation}
and $\alpha=\dfrac{C'\,X^1}{2\,C}+F_{,yy}$ where $X^1(t,x,y,z)$, $A(x)$, $C(x)$, $\Phi(t)$ and $\Omega(t)$ are arbitrary functions while the function $F(y,z)$ satisfy the relation $F_{,yy}(y,z)+F_{,zz}(y,z)\,=\,0$.

\textbf{Case (IX):} In this case we take $B(x)\,\neq\,0$,
$C(x)\,\neq\,0$ and $A(x)\,=\,0$. This case gives the following
CRCs:

\begin{equation}  \label{u41-CRC-IX}
\left\{
  \begin{array}{ll}
    X^0\,=\,X^0(t,x,y,z),\\
\\
X^1\,=\,\sqrt{\dfrac{C(x)}{B(x)}}\,\Bigg[a_1+a_2\,y+a_3\,z+a_4\,\left(y^2+z^2-\left[\int\sqrt{\dfrac{B(x)}{C(x)}}\,dx\right]^2\right)\\
\\
\,\,\,\,\,\,\,\,\,\,\,\,\,\,\,\,\,\,\,\,\,\,\,\,\,\,\,\,\,\,\,\,\,\,\,\,\,\,\,\,\,\,\,\,\,\,
\,\,\,\,\,\,\,\,\,\,\,\,\,\,\,\,\,\,\,\,\,\,\,\,\,\,\,\,\,\,\,\,\,\,\,\,\,\,\,\,\,\,\,\,\,\,
+2\,\left(a_5+a_6\,y+a_7\,z\right)\,\int\sqrt{\dfrac{B(x)}{C(x)}}\,dx\Bigg],\\
\\
X^2\,=a_8+2\,a_5\,y+a_9\,z+a_6\,\left(y^2-z^2-\left[\int\sqrt{\dfrac{B(x)}{C(x)}}\,dx\right]^2\right)\\
\\
\,\,\,\,\,\,\,\,\,\,\,\,\,\,\,\,\,\,\,\,\,\,\,\,\,\,\,\,\,\,\,\,\,\,\,\,\,\,\,\,\,\,\,\,\,\,
\,\,\,\,\,\,\,\,\,\,\,\,\,\,\,\,\,\,\,\,\,\,\,\,\,\,\,\,\,\,\,\,\,\,\,\,\,
+2\,a_7\,y\,z-\left(a_2+2\,a_4\,y\right)\,\int\sqrt{\dfrac{B(x)}{C(x)}}\,dx,\\
\\
X^3\,=a_{10}-a_9\,y+2\,a_5\,z+a_7\,\left(z^2-y^2-\left[\int\sqrt{\dfrac{B(x)}{C(x)}}\,dx\right]^2\right)\\
\\
\,\,\,\,\,\,\,\,\,\,\,\,\,\,\,\,\,\,\,\,\,\,\,\,\,\,\,\,\,\,\,\,\,\,\,\,\,\,\,\,\,\,\,\,\,\,
\,\,\,\,\,\,\,\,\,\,\,\,\,\,\,\,\,\,\,\,\,\,\,\,\,\,\,\,\,\,\,\,\,\,\,\,\,
+2\,a_6\,y\,z-\left(a_3+2\,a_4\,z\right)\,\int\sqrt{\dfrac{B(x)}{C(x)}}\,dx,
  \end{array}
\right.
\end{equation}
and
\begin{equation}  \label{u41-CRC-IX-0}
%\left\{
  \begin{array}{ll}
    \alpha\,=\,\dfrac{\left(a_1+a_2\,y+a_3\,z\right)\,C'(x)}{2\,\sqrt{B(x)\,C(x)}}\\
\\
\,\,\,\,\,\,\,\,\,\,\,\,\,\,\,\,\,\,\,\,\,\,
+a_4\,\left[\dfrac{C'(x)}{2\,\sqrt{B(x)\,C(x)}}\,\left(y^2+z^2-\left[\int\sqrt{\dfrac{B(x)}{C(x)}}\,dx\right]^2\right)-2\,\int\sqrt{\dfrac{B(x)}{C(x)}}\,dx\right],\\
\\
\,\,\,\,\,\,\,\,\,\,\,\,\,\,\,\,\,\,\,\,\,\,
+\left(a_5+a_6\,y+a_7\,z\right)\,\left[2+\dfrac{C'(x)}{2\,\sqrt{B(x)\,C(x)}}\int\sqrt{\dfrac{B(x)}{C(x)}}\,dx\right]
  \end{array}
%\right.
\end{equation}
where $B(x)$ and $C(x)$ are arbitrary functions, $a_i$, $i=1,...,10$
are arbitrary constants.

%%%%%%%%%%%%%%%%%%%%%%%%%%%%%%%%%%%%%%%%%%%%%%%
\section{Conclusion}
%%%%%%%%%%%%%%%%%%%%%%%%%%%%%%%%%%%%%%%%%%%%%%%
In this paper a plane symmetric static spacetime is classified
according to its conformal Ricci collineations by taking the
conformal factor $\alpha$ to be a general function of the spacetime
coordinates. The explicit forms of conformal Ricci vectors along
with constrains on the components of the Ricci tensor are obtained.
When the Ricci tensor is non-degenerate, we have obtained finite
number of CRCs while the Ricci tensor is degenerate, it is concluded
that the Lie algebras of the CRCs is not always finite. It is worth
note that, the numbers of all possible cases of finite and infinite
dimensional Lie algebras of CRCs for the plane symmetric static
spacetime are \textit{ten}. In the non-degenerate case there are
\textit{three} CRCs. Since we obtained highly non-linear
differential constraints on the components of Ricci tensor, so we
have not been able to get exact form of the metric except in case
(V) in sect. 4, where we obtained the spacetime metric which admit
infinite dimensional CRC.

%%%%%%%%%%%%%%%%%%%%%%%%%%%%%%%%%%%%%%%

\end{document}